**Improving Latency with Active Queue Management (AQM) During COVID-19**
By Allen Flickinger, Carl Klatsky, Atahualpa Ledesma, Jason Livingood, Sebnem Ozer
30 July 2021

1. Abstract

During the COVID-19 pandemic the Comcast network performed well in response to unprecedented changes in Internet usage[1][2][3][4]. Video communications applications such as Zoom, WebEx, and FaceTime have exploded in popularity – applications that happen to be sensitive to network latency. However, in today's typical networks – such as a home network – those applications often degrade if they are sharing a network link with other network traffic. This is a problem caused by a network design flaw often described using the term 'buffer bloat'. Several years ago, Comcast helped to fund research & development in the technical community into new Active Queue Management (AQM) techniques to eliminate this issue and this was later built into Data Over Cable Service Interface Specification (DOCSIS) standards[5][6][7].

Just prior to the pandemic, Comcast also deployed a large-scale network performance measurement system that included a latency under load test. In addition, Comcast happened to deploy two otherwise identical cable modems; one with upstream AQM enabled, and the other without. This fortuitous confluence of events has enabled Comcast to perform a comparative analysis of the differences between cable modem gateways using AQM with those that lack that enhancement, at a unique level of scale across many months of time and millions of devices. These deployments were completed during a period of unprecedented network bandwidth demand growth due to work and school from home during the pandemic.

The data reveals significantly better upstream latency under load performance when the Proportional Integral Controller Enhanced (DOCSIS-PIE) AQM is used. For the device with AQM, the large majority of the latency under load tests resulted in 15-30 milliseconds of latency. In comparison, the device without AQM averaged roughly 250 milliseconds of latency – between 8-16 times higher – a highly significant difference to the end user quality of experience (QoE). As a result, users of this XB6 device with AQM enabled will have a far superior experience when using all Internet applications. Delay is sometimes thought to affect primarily latency-sensitive applications such as video conferencing and online gaming, but consistent low delay provides a better user experience for any Internet application where the user is waiting to see the result of some operation, such as web browsing, stock quotes, weather forecasts, maps,

---

[1] See 2020 network performance report at https://corporate.comcast.com/press/releases/comcast-2020-network-performance-data
[2] See report on Pandemic Network Performance from the Broadband Internet Technical Advisory Group (BITAG) at https://www.bitag.org/2020-pandemic-network-performance.php
[3] See report from an Internet Architecture Board workshop on COVID at https://www.rfc-editor.org/rfc/rfc9075.html
[4] See Comcast-specific reports at https://corporate.comcast.com/covid-19/network/may-20-2020 and https://corporate.comcast.com/stories/covid-19-network-report-smart-network-speed-and-stability
[5] https://community.cablelabs.com/wiki/plugins/servlet/cablelabs/alfresco/download?id=3a751ba9-593c-464a-bc91-3675e1824e30
[6] https://tools.ietf.org/id/draft-white-aqm-docsis-pie-00.html,
[7] https://www.res.cablelabs.com/wp-content/uploads/2019/02/28094021/DOCSIS-AQM_May2014.pdf

**Improving Latency with Active Queue Management (AQM) During COVID-19**

driving directions, etc. Even video streaming applications benefit from lower network delay (latency) because they need less initial buffering time before a video can begin playing. These large-scale measurement comparisons should provide additional data to justify accelerated deployment of AQM in DOCSIS and other Internet Service Provider (ISP) networks, particularly in the equipment installed in customer homes, including user-purchased home network equipment.

## 2. Introduction

In 2020, the COVID-19 pandemic emerged as an extremely disruptive global event, leading to seismic changes in everyone's day-to-day lives. In particular, the pandemic prompted a sudden, massive shift to home-based work & school. This led to unprecedented increases in network usage for ISPs across the world, including significant new demands on home networks and local access network segments.

The key application to emerge and achieve mainstream adoption was real-time video conferencing, such as Zoom, which on the Comcast network increased in volume by 316% between February and October 2020[8]. As opposed to other types of applications, such as overnight online backups or background downloads of video game updates, the quality of the end user experience is highly delay-sensitive – meaning it is significantly affected by any network latency. Even simple web browsing is improved by lower network delays, though in this case users have been trained over the last 25 years to simply to be more patient when web pages are sometimes slow to load for no apparent reason.

While it has always been the case that real-time communications traffic is latency-sensitive, widespread adoption of video conferencing in parallel with significantly increased competing LAN traffic in a home during the COVID-19 pandemic has created a perfect storm of using latency-sensitive applications while other Internet traffic is sharing the home network. This can lead users to experience a degraded video conferencing experience – because of "latency under load" or higher "working latency" – during times when more than one kind of traffic is sharing the network.

High latency under load and buffer bloat[9], are issues that the networking community has been aware of for many years and Comcast and other ISPs and software developers have been working on solutions to this challenge. In the case of DOCSIS networks, the solution has been to shift from legacy Tail-Drop First-In First-Out (FIFO) queues and configurable buffer control in DOCSIS 3.0, to Active Queue Management (AQM)[10] – a feature that is implemented DOCSIS and many other networks and software systems. Enabling AQM on a network interface that suffers from poor responsiveness under normal working conditions improves the performance of all user-facing network applications, and especially real-time, latency-sensitive applications such

---

[8] Based on analysis of aggregate network changes from NetFlow, interconnection traffic exchange, and similar data sources.
[9] See Buffer Bloat article, "Buffer Bloat: Dark Buffers in the Internet" by Jim Gettys in the November 29, 2011 issue of Communications of the ACM at https://queue.acm.org/detail.cfm?id=2071893
[10] RFC 7567 at https://www.rfc-editor.org/rfc/rfc7567.html is a great background to why AQM is necessary, so is an excellent starting point for readers learning about the AQM and the context into which it fits in the history of the Internet.

**Improving Latency with Active Queue Management (AQM) During COVID-19**

as video conferencing, without any negative effects on overall throughput or on delay-tolerant applications.

This paper explains how and where AQM was deployed in the Comcast network. It also compares latency under load performance measurements for those devices that utilized AQM to those without AQM support.

3.  **Developments in 2020**

In early 2020 Comcast implemented a new cable modem-based Quality of Experience (QoE) measurement system. This QoE measurement system, notably for this report, included a latency under load measurement[11]. The measurement system was deployed in early 2020 prior to the pandemic [12] [13]. It consisted of millions of devices able to run performance tests, thus creating a baseline of latency under load measurements prior to the pandemic.

In addition, of the range of Comcast cable modem gateways and retail cable modems, the Comcast XB6 device has two distinct model variants – one with upstream AQM enabled during the measurement period in 2020 and one without. This provided a fortuitous opportunity for a uniquely high scale side-by-side performance comparison for a specific model of Comcast home gateway at a time of high demand for latency-sensitive applications. Comcast also in late 2020 deployed and optimized downstream AQM on all Arris E6000 Cable Modem Termination Systems (CMTSes)[14].

Looking at the events and changes of the year 2020, it is clear that latency-sensitive applications such as video conferencing are here to stay. As well, implementing AQM and other low latency improvements can very positively impact end user QoE. Such improvements can also be deployed independently by ISPs and equipment manufacturers, with no need to coordinate with application developers[15]. This should provide motivation for network operators to deploy AQM and other low latency improvements and we believe that latency under load performance will emerge as a competitive differentiator – perhaps eventually on equal footing as connection speed in the minds of consumers once they learn how critical latency under load is to their perception of performance.

---

[11] The test uses the open Realtime Response Under Load (RRUL) specification per
https://www.bufferbloat.net/projects/bloat/wiki/RRUL_Spec/
[12] See NetForecast's independent audit of the design of this measurement system at
https://www.netforecast.com/netforecast-design-audit-report-of-comcasts-network-performance-measurement-system/
[13] See NetForecast's independent assessment of performance results during the pandemic at
https://www.netforecast.com/netforecasts-report-on-comcasts-network-performance-measurement-system-results-data/
[14] Deployment occurred between August 12, 2020, and September 16, 2020.
[15] This is a good demonstration of the fast-moving and loosely coupled nature of the Internet, where different parts of the Internet ecosystem do not need to coordinate their actions in advance.

**Improving Latency with Active Queue Management (AQM) During COVID-19**

**4. Why Does Latency (Delay) Matter?**

Latency is the delay that can occur on a network in sending or receiving packets and this can be caused by a wide range of factors[16], from the distance traversed on the network or the way that applications and protocols are designed, to load on a home or access network, to the way that operating systems, application platforms, datacenters, network interface cards, broadband gateways, network switches, and network routers function.

Different types or classes of applications have widely differing sensitivities to latency. Generally though, applications can be grouped into two major latency categories: delay *sensitive* and delay *tolerant*. Delay sensitive applications include video conferencing, voice over IP (VoIP), online video games, web browsing, desktop remote control, live streaming, instant messaging, streaming video. Delay sensitive applications also include anything where a human user is waiting to see the result of some operation when interacting with a screen or device, such as web browsing, stock quotes, weather forecasts, maps, driving directions, etc. Delay tolerant applications include cloud-based hard drive backups, large game, or other background file downloads such as operating system updates, and peer to peer file sharing (P2P).

Most users can relate to the experience of using a delay sensitive application and then encountering latency. For example, participating in a video conference call trying to sing or play music together[17] or having audio or video intermittently freeze up, or playing an online game and a sudden increase in latency leading to a slow play reaction and lost game.

What users experience is their so-called Quality of Experience (QoE); a subjective measure of the customer's level of satisfaction when using the Internet and various applications. Traditionally, the Mean Opinion Score (MOS) is used as an arithmetic mean of all the individual QoE scores. With advanced applications of Machine Learning, new systems have been developed to predict the customer's QoE based on the measurable and objective QoE metrics and other factors including human perception, past experiences, expectations, and service price. Hence, QoE is a multidisciplinary concept involving engineering, cognitive neuroscience, economics, statistics, and social psychology[18].

For instance, human reaction time is a combination of different components of mental processing including sensory perception, receipt of input into our consciousness, context applied to the input and decision made based on processing output[19]. Recent research that mere milliseconds of delay

---

[16] See "Reducing Internet Latency: A Survey of Techniques and their Merits" at https://ieeexplore.ieee.org/document/6967689

[17] Playing music together over the Internet poses some interesting technical challenges. See https://www.npr.org/2020/11/21/937043051/musicians-turn-to-new-software-to-play-together-online and https://www.jacktrip.org/ for more about this.

[18] See the QoE section at "From OBI and SNR to Ookla and Netflix: How Network Impairments affect Customer Perceptions: The role of Leading and Lagging Indicators as We Evolve HFC to FTTP"; by Sebnem Ozer, Venk Mutalik, A. Vieira, and J. Chrostowski in the SCTE Cable-Tec Expo 2015

[19] See https://www.pubnub.com/blog/how-fast-is-realtime-human-perception-and-technology/ for insights on how fast a human can process input.

**Improving Latency with Active Queue Management (AQM) During COVID-19**

is perceptible and thus matters to the user's QoE[20] [21] [22] [23]. While mean and median latency is important, for latency sensitive applications even minor increases at the 99th percentile can affect a user's QoE. In addition, it is also important for applications to have not just low latency but also *consistently low* latency. Note however, that it is a rare networked application that is not improved by consistent lower network delay; improving latency improves more than just video conferencing and online gaming.

5. **Types of Cable Modem Devices**

As explained earlier, latency under load can occur at any place along a network path, typically at the entrance of the most constrained link in the path (a.k.a. the bottleneck link). By design, the likely place a user may experience latency under load is in their Internet access device because this is the device where their advertised speed is provisioned. In the Comcast network, like many other ISP networks, the wired and wireless (WiFi) local area network (LAN) functions and routing functions are most often performed by one integrated gateway device. In the Comcast network, that is the cable modem gateway, though customers can instead install their own modem and a separate home router and/or WiFi access point(s) customer premises equipment (CPE).

Comcast's Internet service uses the Data Over Cable Service Interface Specifications (DOCSIS) standard as the data link layer protocol. This enables a network connection from the home's cable modem (CM) to Comcast's Cable Modem Termination System (CMTS). This provides the connection between the customer's home network and the Internet.

The DOCSIS suite of specifications are publicly available[24], allowing various Original Equipment Manufacturers (OEMs) to design & develop cable modems. Customers have a choice for their cable modem; they can purchase one themselves (e.g., from Best Buy or Amazon) or they can lease one from their ISP (e.g., Comcast). When a customer purchases their own cable modem, they are responsible for administering it, updating the software, configuring it, replacing it if it fails, and so on. These modems are generally referred to as Consumer Owned And Managed (COAM) devices.

An important distinction between leased and COAM modems is support for the operating firmware. For COAM devices, the modem's operating firmware is provided by the modem's manufacturer, who controls the feature set, bug fixes, and firmware release schedule (to the extent that there even are any post-sale software updates). In contrast, leased devices are generally remotely administered, configured, and regularly updated by the ISP, which can bring a range of practical benefits, especially relating to performance and security. The decision to

---

[20] See "Neuronal Processing: How fast is the speed of thought?", by Tovée, Martin in Current Biology: CB. 4, 1995
[21] See "Detecting meaning in RSVP at 13 ms per picture", by Mary C. Potter, Brad Wyble, Carl Erick Hagmann & Emily S. McCourt, in Attention, Perception, & Psychophysics, volume 76, 2014
[22] See "Latency and Cybersickness: Impact, Causes, and Measures. A Review" at https://www.frontiersin.org/articles/10.3389/frvir.2020.582204/full
[23] See Appendix A.1 of "Problem Statement: Transport Support for Augmented and Virtual Reality Applications" at https://datatracker.ietf.org/doc/html/draft-han-iccrg-arvr-transport-problem-01#appendix-A.1
[24] Please refer to footnote 1.

**Improving Latency with Active Queue Management (AQM) During COVID-19**

include support for AQM for COAM devices is in the hands of the device manufacturer, whereas in the case of leased devices the operating firmware and configuration of that firmware is controlled by Comcast.

In earlier generations of Comcast's leased devices, the manufacturer still played a prominent role in the development of the operating firmware. But this approach did not allow much flexibility, software release schedules were heavily influenced by the manufacturers, and those release cycles were quite long in duration.

As a result, starting around 2011, Comcast and other cable-based ISPs began to take a more active role in the development of cable modem software. Many cable ISPs have embraced a more open source-based methodology for cable modem firmware development which has led to the current generation of DOCSIS 3.0 and DOCSIS 3.1 cable modem software to being based on the open source "Reference Design Kit – Broadband" (RDK-B)[25] software stack. As the RDK website explains[26], *"The RDK is a standardized software stack with localization plugins created to accelerate the deployment of next-gen broadband products and services by broadband service providers."*

With the adoption of the RDK-B software stack, Comcast and other cable ISPs have become the software developers for the cable modems, running on hardware supplied by the Original Equipment Manufacturer (OEM) partners. The cable ISP development team makes the decisions on which software features are included in each operating firmware release and has fine-grained control over which devices to deploy updates, when, and in what numbers. RDK also makes it possible for a range of parties in the ecosystem to develop and release software, but even if a given software module is developed by the OEM, the cable ISP determines if and when it is included in a future software release and when it is deployed to customer devices.

In the Comcast network, RDK-B devices include the Xfinity XB3, XB6 and XB7 devices[27], with hardware produced by a range of OEMs and System on Chip (SOC) suppliers. The XB3 is based on DOCSIS 3.0, while XB6 and XB7 are based on DOCSIS 3.1. As a result of the new capability to manage and update software that is made possible by RDK-B, as well as the deployment of DOCSIS 3.1, Comcast was able to introduce AQM into the software of one variant of the XB6 gateway beginning in 2017.

## 6. Latency Under Load Research & Development History at Comcast

Comcast engineers Rich Woundy and Jason Livingood started discussing the concept of latency under load with Jim Gettys, who coined the term "buffer bloat" in 2010. This later expanded to include Dave Täht, one of Gettys' colleagues. In 2011, Gettys began an industry-wide roadshow

---

[25] See the RDK website at https://rdkcentral.com/
[26] See RDK's web page for developers at https://developer.rdkcentral.com/broadband/documentation/getting_started/
[27] See overview of the Xfinity gateway series and specific models at https://www.xfinity.com/support/articles/broadband-gateways-userguides

**Improving Latency with Active Queue Management (AQM) During COVID-19**

to explain the problem and possible solutions[28] which was quite influential and helped motivate the technical community to study this issue in greater detail.

The following year, in 2012, Comcast entered into a consulting agreement with Dave Täht to help underwrite development of an implementation of the "CoDel" (Controlled Delay) AQM algorithm[29]. In parallel, several CableLabs engineers including Greg White and Joey Padden started work to assess the impact of the CoDel AQM in DOCSIS networks, noting a 2.7-point improvement[30] in VoIP Mean Opinion Score (MOS)[31] points and markedly better web page loading performance. That same year, Comcast provided a letter of support for a buffer bloat research grant that Dr. Kathleen Nichols had applied for with the U.S. Department of Energy's Office of Science[32]. Dr. Nichols eventually received two grants for this work in 2013-2014[33].

Also in 2012, flowing from his work on CoDel, Dave Täht started work to develop a latency under load benchmarking test, which Täht called the Realtime Response Under Load (RRUL) test[34]. He developed this in partnership with Toke Høiland-Jørgensen from Roskilde University. Täht also began to work to implement CoDel in home gateway devices using the OpenWrt open-source home gateway software, eventually resulting in a fork of OpenWrt code called CeroWrt[35] [36]. In addition, Eric Duzamet developed an implementation called "fq_codel" for Flow Queuing Controlled Delay. Implementation of these AQMs was also added to the Linux kernel and the mainstream OpenWrt source code and downstream products including Tomato, dd-wrt, and Ubiquity's software.

In late 2012 and early 2013, CableLabs engineers Greg White and Dan Rice (now at Comcast) conducted a study of several AQMs as candidates for inclusion in future DOCSIS specifications[37]. Comcast also conducted demonstrations at the CableLabs Winter Conference in 2013.

---

[28] One of the earliest was in a Google talk in April 2011, see https://www.youtube.com/watch?v=qbIozKVz73g followed by a talk at NANOG-52 in June 2011, see slides at https://archive.nanog.org/meetings/nanog52/presentations/Tuesday/Gettys-NANOG11a.pdf

[29] The CoDel algorithm itself was conceptualized by Dr. Kathleen Nichols and Van Jacobson. Both are key figures in the history of TCP congestion control. For more information see https://en.wikipedia.org/wiki/Van_Jacobson and https://en.wikipedia.org/wiki/Kathleen_Nichols. Dr. Nichols later worked with CableLabs on AQM in DOCSIS 3.0, while Van Jacobson was hired by Google and was part of developing the new BBR and QUIC protocols – see https://www.networkworld.com/article/3218084/how-google-is-speeding-up-the-internet.html.

[30] See https://www-res.cablelabs.com/wp-content/uploads/2019/02/28094033/PreliminaryStudyOfCoDelAQM_DOCSIS-Network.pdf

[31] The Mean Opinion Score is a 5-point quality scale for voice communications. See https://en.wikipedia.org/wiki/Mean_opinion_score for more info.

[32] Small Business Innovation Research/Small Business Technology Transfer 2013 Phase 1 Funding Opportunity (DE-FOA-0000760)

[33] See grant awards at https://www.sbir.gov/sbirsearch/detail/687053 and https://www.sbir.gov/sbirsearch/detail/407679

[34] See RRUL specification at https://www.bufferbloat.net/projects/bloat/wiki/RRUL_Spec/

[35] See CeroWrt background at https://www.bufferbloat.net/projects/cerowrt/wiki/Historical_Documents/

[36] The first CeroWrt release was announced May 14, 2012, in https://lists.bufferbloat.net/pipermail/cerowrt-devel/2012-May/000233.html

[37] Several were tested, such as CoDel, SFQ-CoDel, and PIE. See paper at https://www-res.cablelabs.com/wp-content/uploads/2019/02/28094033/Active_Queue_Management_Algorithms_DOCSIS_3_0.pdf

**Improving Latency with Active Queue Management (AQM) During COVID-19**

In 2013, work had progressed to a point where the CoDel AQM could be demonstrated in DOCSIS networks and so Jason Livingood, Chris Griffiths and other Comcast engineers conducted demonstrations at IETF-86 in March 2013[38], along with Dave Täht as well. That same month, in advance of the IETF meeting, this team also conducted a series of executive demos with and without AQM, using a CeroWrt-based router connected to a cable modem, running the Quake online game, running the Chrome browser performance benchmark test, and running a video conference session. This implementation and associated testing demonstrated meaningful quantitative performance improvements, such as probability of gaming latency under network load of roughly 10 milliseconds with AQM compared to 1 second without AQM (100x worse).

In 2014, CableLabs included PIE as the official AQM in the DOCSIS 3.1 specifications as a mandatory feature for the CMTS and cable modems[39] (PIE was also added as an optional feature for earlier generation DOCSIS 3.0 cable modems). At the same time, CableLabs required that DOCSIS 3.1 CMTS equipment support AQM as well but left the algorithm choice to the vendor. That same year, Comcast provided further financial support for Dave Täht to perform research and development to improve buffer bloat in WiFi network links and equipment, a project he called "Make WiFi Fast" that focused on achieving "airtime fairness" and improved latency under load on the WiFi link.

In 2015, led by Carl Klatsky, with data analytics support from Yana Kane-Esrig, engineering support from Andrew Mulhern and Drew Taylor, as well as guidance from Chae Chung, Hal Kim, and Jason Livingood, Comcast performed limited field trials of a low latency service, using a commonly deployed modem, the Technicolor TC8305C, which supported adjustments to the modem's buffer size through configuration updates. Adjusting the modem's buffer size through this "buffer control" feature served as a pre-cursor technique for latency mitigation prior to AQM implementations being added to the modems. This field trial validated real-world performance differences using latency mitigation techniques in a DOCSIS network and cable modem buffer control configurations were eventually deployed into the Comcast network in 2019.

Comcast has also funded research & development via the Comcast Innovation Fund[40] that was focused on additional improvements in latency under load performance, as summarized below.
- In 2015, via two grants to Teklibre for work on open-source development of buffer bloat mitigation measures as well as mitigations specifically for WiFi.
- In 2017, a WiFi open-source development grant was extended to help Teklibre work on addressing WiFi multicast buffer bloat issues.
- In 2019, a grant was made to support Bob Briscoe's continued work on Low Latency Low Loss Scalable throughput (L4S)[41] following development of the initial CableLabs

---

[38] See video at https://www.youtube.com/watch?v=NuHYOu4aAqg
[39] See CableLabs blog at https://www.cablelabs.com/how-docsis-3-1-reduces-latency-with-active-queue-management and paper at https://www.cablelabs.com/wp-content/uploads/2014/06/DOCSIS-AQM_May2014.pdf
[40] https://innovationfund.comcast.com/
[41] See https://datatracker.ietf.org/doc/html/draft-ietf-tsvwg-l4s-arch and https://community.cablelabs.com/wiki/plugins/servlet/cablelabs/alfresco/download?id=0affb960-25f4-44f9-b747-74ad8555fd06

**Improving Latency with Active Queue Management (AQM) During COVID-19**

   specifications of Low Latency DOCSIS[42], which supports L4S. He in turn funded other open source developers[43].
- In 2020, Teklibre received an additional grant to work on accelerating AQM adoption in the real world.
- In 2021, grants have been awarded to Bob Briscoe to continue work on L4S and to Teklibre to work on WiFi latency under load when multiple users are simultaneously demanding over 100Mbps on the wireless LAN.

In summary, the deployment of AQM into the Comcast DOCSIS network did not happen easily or by chance. The first step was to get the technical community to recognize that the Internet had a problem and to focus on likely solutions, and that critical early work was done brilliantly and passionately by Jim Gettys. This led to theorizing about, developing, and testing potential solutions over the course of many years of work and experimentation. It would not have been successful without the deep thinking and research of a handful of highly knowledgeable, committed, and passionate engineers and developers, of whom we have mentioned just a few. The success of this effort also is due substantially to work within the open source software community, the involvement of many Internet end users that were willing to do their part to help test, to collaborative development of CableLabs DOCSIS standards (spanning CableLabs staff and contractors, ISPs, and vendors), and finally to implementation support by both DOCSIS vendors and the RDK-B software community.

## 7. AQM in DOCSIS

In DOCSIS 3.0 there was an optional "buffer control" parameter for the cable modem and CMTS that could be used but it was difficult to configure and was a static buffer setting rather than a dynamic AQM[44]. As configured in the Comcast network, when using buffer control with DOCSIS 3.0 devices, this would typically result in a maximum upstream latency of roughly 250 milliseconds.

But buffer control is not as flexible as AQM, which has the advantage of being able to absorb temporary traffic bursts while supporting smaller average latency and maintaining high throughput. After DOCSIS 3.0 was initially developed, a specification change was issued that *optionally* added the PIE AQM to DOCSIS 3.0 in 2014[45] for cable modems, but by that time development focus had largely shifted to DOCSIS 3.1. This same PIE AQM became a

---

[42] See https://community.cablelabs.com/wiki/plugins/servlet/cablelabs/alfresco/download?id=0affb960-25f4-44f9-b747-74ad8555fd06

[43] Ilpo Järvinen and Asad Ahmed

[44] See "Right Sizing Cable Modem Buffers for Maximum Application Performance" https://www.nctatechnicalpapers.com/Paper/2011/2011-right-sizing-cable-modem-buffers-for-maximum-application-performance/download#:~:text=The%20new%20feature%20is%20referred,well%20as%20a%20target%20size and "DOCSIS Best Practices and Guidelines: Cable Modem Buffer Control" at https://community.cablelabs.com/wiki/plugins/servlet/cablelabs/alfresco/download?id=a4c077e2-cc55-4a42-a557-32f92e637266

[45] See base specification at https://www.cablelabs.com/specifications/CM-SP-MULPIv3.0. Page 791 shows the "Engineering Changes for CM-SP-MULPIv3.0-I24-140403", including "Active Queue Management Requirements" dated 2/26/2014. The actual text for the PIE AQM requirements for the cable modem is in Annex M, pages 719-722. It is referenced by the general AQM requirements in section 7.6.2, pages 237-238.

**Improving Latency with Active Queue Management (AQM) During COVID-19**

mandatory feature in the subsequent DOCSIS 3.1 cable modem standards. AQM is also a mandatory feature of the DOCSIS 3.1 Cable Modem Termination System (CMTS) standards, though the choice of AQM algorithm is left to the vendor[46].

When implemented in the CMTS, AQM affects the downstream queue from the CMTS to the cable modem. In contrast, when implemented in the cable modem, the AQM affects the upstream queue from the cable modem to the CMTS. Each implementation will bring a benefit to the end user QoE since latency under load can become an issue in both the upstream and downstream directions. Optimally, AQM is deployed on both CMTS and cable modem. Configuration of the CMTS AQM parameters can have a significant effect on downstream latency as described in a recent paper by Sebnem Ozer, et al.[47]

In addition, CableLabs has developed and tested a new DOCSIS 3.1 extension that reduces medium access (MAC) delays in the upstream and supports the IETF Non-Queue Building (NQB[48]) and Low Latency, Low Loss, Scalable Throughput (L4S) services in either or both directions to create the Low Latency DOCSIS specification[49].

Of the Comcast leased cable modem gateway devices, RDK-B software is running on XB3, XB6 and XB7[50]. AQM development has been focused on DOCSIS 3.1-based devices, such as the greater than 10 million XB6 devices and the latest generation XB7 device. As well, customer-owned DOCSIS 3.1 cable modem devices also should support AQM.

8. **Devices & Software Under Study**

For the XB6, there are two different hardware manufacturers, which means two variants of XB6. Those are the Technicolor CGM4140COM and Arris (now CommScope) TG3482G. The XB6 is the focus of this analysis because one of these variants – the Technicolor device - had AQM enabled from its introduction in October 2017, whereas the other – the CommScope device - did not during the study period [51]. This is simply a result of differences in the chipsets used and the readiness and ability to add AQM code[52].

There were roughly 4.3M of the AQM-enabled Technicolor devices deployed as of December 2020, in comparison to 2M non-AQM-enabled CommScope devices. In the early part of 2021, AQM was enabled on the remaining variant of XB6.

---

[46] While AQM can be enabled in the CMTS, support varies by implementation. Different CMTS models have adopted different AQM functionalities and in some cases proprietary algorithms.
[47] Approaches to Latency Management: Combining Hopby-Hop and End-to-End Networking. Ozer, Klatsky, Rice, Chrostowski. SCTE 2020. https://www.nctatechnicalpapers.com/Paper/2020/2020-approaches-to-latency-management-combining-hopby-hop-and-end-to-end-networking
[48] See "A Non-Queue-Building Per-Hop Behavior (NQB PHB) for Differentiated Services" at https://datatracker.ietf.org/doc/draft-ietf-tsvwg-nqb/
[49] See https://www.cablelabs.com/technologies/low-latency-docsis
[50] See device guide at https://www.xfinity.com/support/articles/broadband-gateways-userguides
[51] These were eventually traced to DOCSIS 3.1 PHY configuration issues, which have subsequently been resolved.
[52] Some DOCSIS 3.1 features such as AQM had to be disabled shortly after deployment began due to performance issues. The cause was eventually traced to DOCSIS 3.1 PHY configuration issues and resolved.

**Improving Latency with Active Queue Management (AQM) During COVID-19**

This unique situation of two variants of the same device, deployed to millions of homes in the United States, one of which has AQM and one that does not, presented the opportunity for an extremely interesting comparative case study in the performance impacts of AQM. This direct comparison is much more compelling than contrasting a newer XB7 with an older XB6, given greatly differing CPU, memory, radio, and network interface specifications. It provided an unparalleled opportunity to directly compare latency under load with and without AQM at a time when latency sensitive video conferencing and other Internet traffic had dramatically increased.

## 9. Latency Testing

Previous latency testing options either introduced too many variables to make any definitive conclusions or did not offer the ability to scale in a way that would offer insight to ongoing network trends. As a result, Comcast uses a custom Network Performance Measurement System that leverages open-source software. This system aims to eliminate as many user device and WiFi-related variables as possible, focusing on those conditions and variables over which Comcast has direct control, while also being able to work at high levels of scale[53]. Figure 1 shows a high-level view of the system with the following components:

- **Measurement Orchestration (orchestration)** maintains a holistic view of the network between the client and the servers and initiates measurements accordingly.
- **Measurement Client (client)** will, when instructed by the orchestration, run upstream and downstream measurements to and from the Measurement Servers.
- **Access Network** is the internal Comcast network, from the DOCSIS portion of the network to which users connect as well as regional and backbone portions of the network.
- **Measurement Server (server)** dedicated hardware servers responding to measurement requests.

*Figure 1: Comcast's Network Performance Measurement System Architecture*

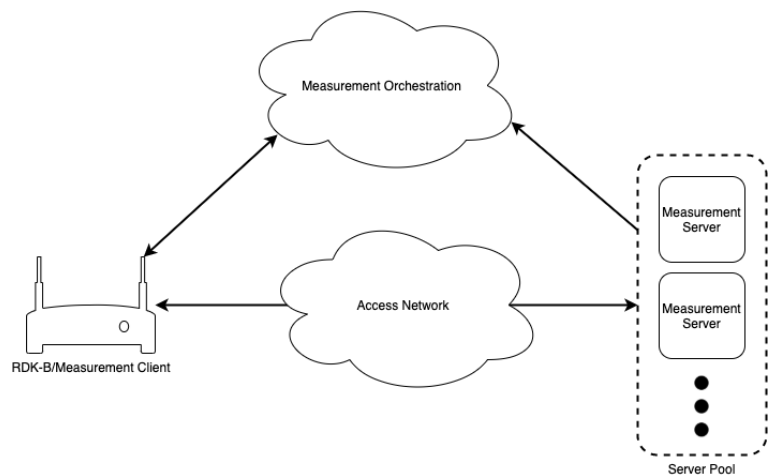

---

[53] The design of the Comcast Network Performance Measurement System was independently reviewed by NetForecast. See their report at https://netforecast.com/wp-content/uploads/Comcast-Design-Audit-Report-NFR5133F.pdf.

**Improving Latency with Active Queue Management (AQM) During COVID-19**

The platform actively introduces test traffic rather than passively measuring user-generated traffic.

The client is built into the RDK-B firmware that is running on the customer premise-based cable modem gateway. The orchestration function takes in measurement requests and determines whether the measurement will have enough available measurement system resources to sustain the peak load and duration of a given test. That resource check is performed on both the client and the server.

If either of those checks fails, the measurement request is rejected, and the test will be tried again later. If the measurement is accepted, the orchestration function will send a request to the client with all the parameters needed to perform the measurement, such as the type of measurement, duration of measurement, and server(s) to use. The orchestration function also instructs the server to reserve those resources for the client and then the client will start its measurement test cycle. Once the measurement test cycle is complete, the client will request that the server reallocate the reserved resources back into the server's available resource pool. The server then communicates any allocation or deallocation of resources back to the orchestration function.

The first measurement implemented at scale was a network capacity measurement, more generally known as a "speed test". Following NetForecast's design audit of the system, they also performed an audit of the results of speed test measurements[54].

Latency is one of the biggest influencers of end user application performance but can vary greatly based on routine network usage (when a network flow is filling a queue on the network path between two hosts). To determine the impact to latency during periods when a data flow is causing queues to be filled to capacity the system uses a measurement that will simultaneously load the network and measure latency. This measurement, aptly named latency under load, utilizes iperf3[55] to load the network in either the downstream or upstream direction between the client and a server. Figure 2 shows a high-level view of this test.

*Figure 2: High level functional test design*

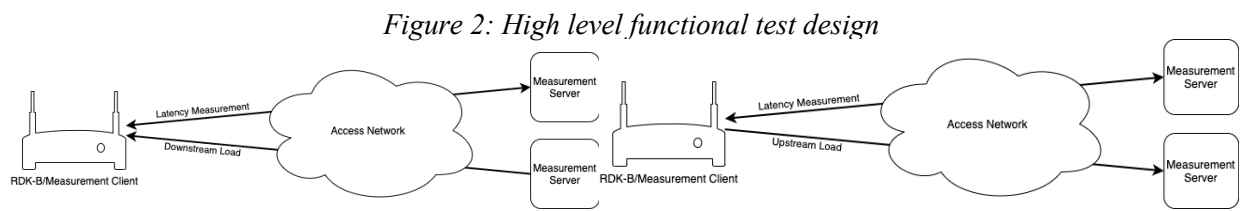

Latency testing begins as the TCP ramp-up of the bandwidth measurement reaches its peak steady state and finishes before the bandwidth measurement connection is closed. The latency measurement sends UDP datagrams back and forth between the client and the server and measures the round-trip latency multiple times over a set interval.

---

[54] See NetForecast report at https://netforecast.com/wp-content/uploads/Comcast-Phase-II-Audit-Report-NFR5134F.pdf
[55] For more information on iperf3, see https://software.es.net/iperf/

**Improving Latency with Active Queue Management (AQM) During COVID-19**

This test uses netperf to measure the delay after every UDP request/response, defined as "UDP_RR"[56]. The duration of the test is configurable and in this case is set at 11 seconds for the downstream measurement and 7 seconds for upstream[57].

**10. Latency Measurement Results**

As explained earlier, for two variants of XB6 cable modem gateway, upstream DOCSIS-PIE AQM was enabled on the CGM4140COM (experiment) variant but was not available on the TG3482G (control) variant during the measurement period. The TG3482G variant used a buffer control configuration that predated AQM in DOCSIS[58]. While measurements were performed throughout the course of 2020, the figures that follow represent over 26,000 tests run using random samples of RDK-B cable modems across the Comcast network in the United States between October 1 and December 31, 2020.

At a high level, when a device had AQM it consistently experienced between 15-30 milliseconds of latency under load. In contrast, without AQM but using the older buffer control configuration, a device would experience significantly higher and more variable of working latency. Those non-AQM devices experienced in many cases 250 milliseconds or higher latency under load. This demonstrates two key points: not only does AQM enable significantly better latency under load performance, but that latency performance is much more uniform and consistently (e.g., relatively lower jitter). Both are excellent outcomes that should improve end user quality of experience for all Internet applications.

Below are two Cumulative Distribution Functions (CDFs), comparing both mean and max latency under load for the two XB6 variants. The CDFs and the bar chart that follows are comprised presents the variability across all devices of a given type and compares the two device variants.

These two CDFs in Figures 3 and 4 show that, whether on average or in the worst case, nearly all devices with AQM will experience 15-30 milliseconds of latency under load. At the same time, without AQM, will experience much higher latency under load. When comparing the mean to the maximum in Figures 3 and 4, respectively, it is also apparent that the devices with AQM perform quite similarly. This suggests that the latency performance is consistently good.

---

[56] See netperf documentation at https://hewlettpackard.github.io/netperf/doc/netperf.html#UDP_005fRR
[57] This period is after TCP ramp-up. The downstream interval is longer as it includes some omit intervals. Please refer to the netperf documentation for additional information
[58] So, when comparing with and without AQM, it is really with AQM or with buffer control.

**Improving Latency with Active Queue Management (AQM) During COVID-19**

*Figure 3 – CDF of Mean Upstream Latency Under Load, Equivalent Devices with and Without AQM*

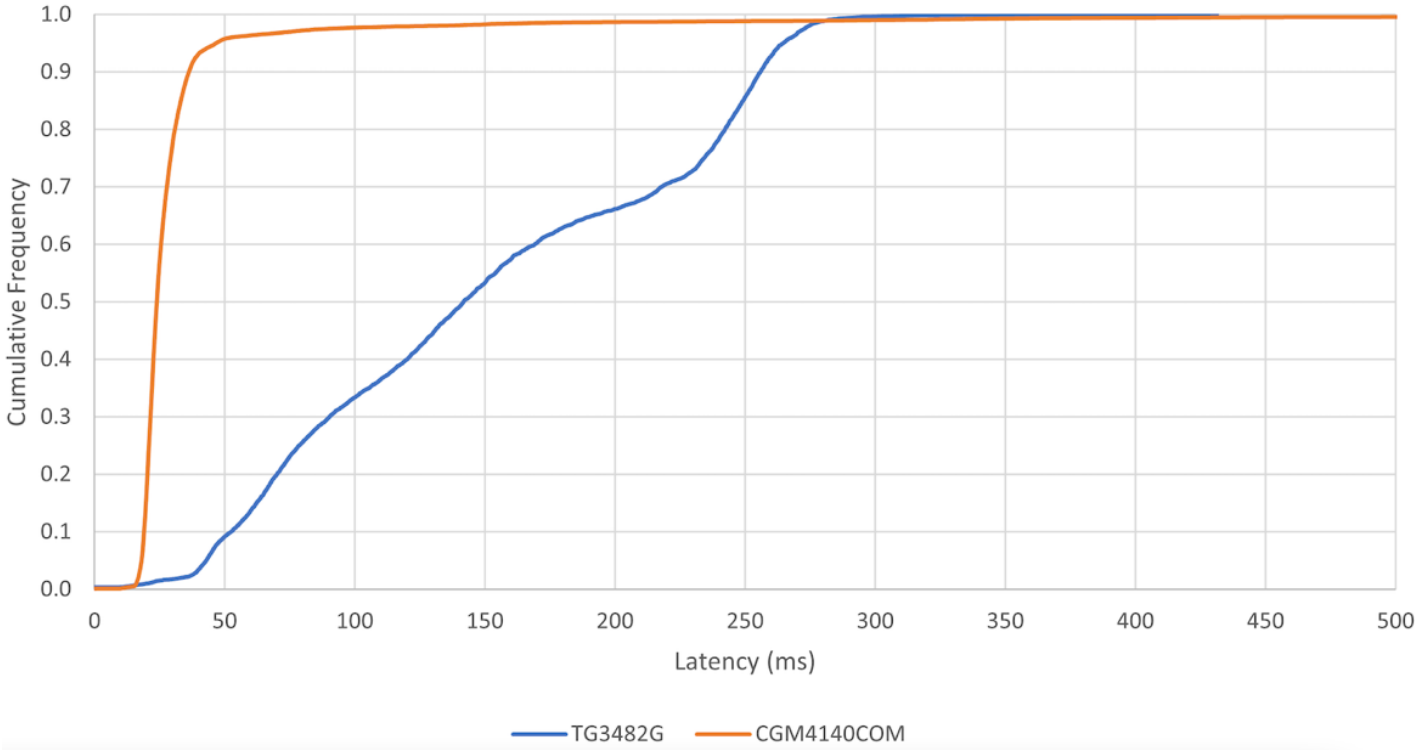

**Improving Latency with Active Queue Management (AQM) During COVID-19**

Figure 4 below shows the CDF of the maximum (worst-case) latency under load performance. The device that lacked AQM, but had an older buffer control configuration, experienced much higher latency compared to its mean in Figure 3, whereas the mean and max for the AQM-enabled device is quite similar. It is also interesting to observe that there was a much longer tail of performance of the non-AQM device, trailing off from a roughly 0.7 cumulative frequency to performance levels that stretch over 1 full second in some cases.

*Figure 4 – CDF of Max Upstream Latency Under Load, Equivalent Devices with and Without AQM*

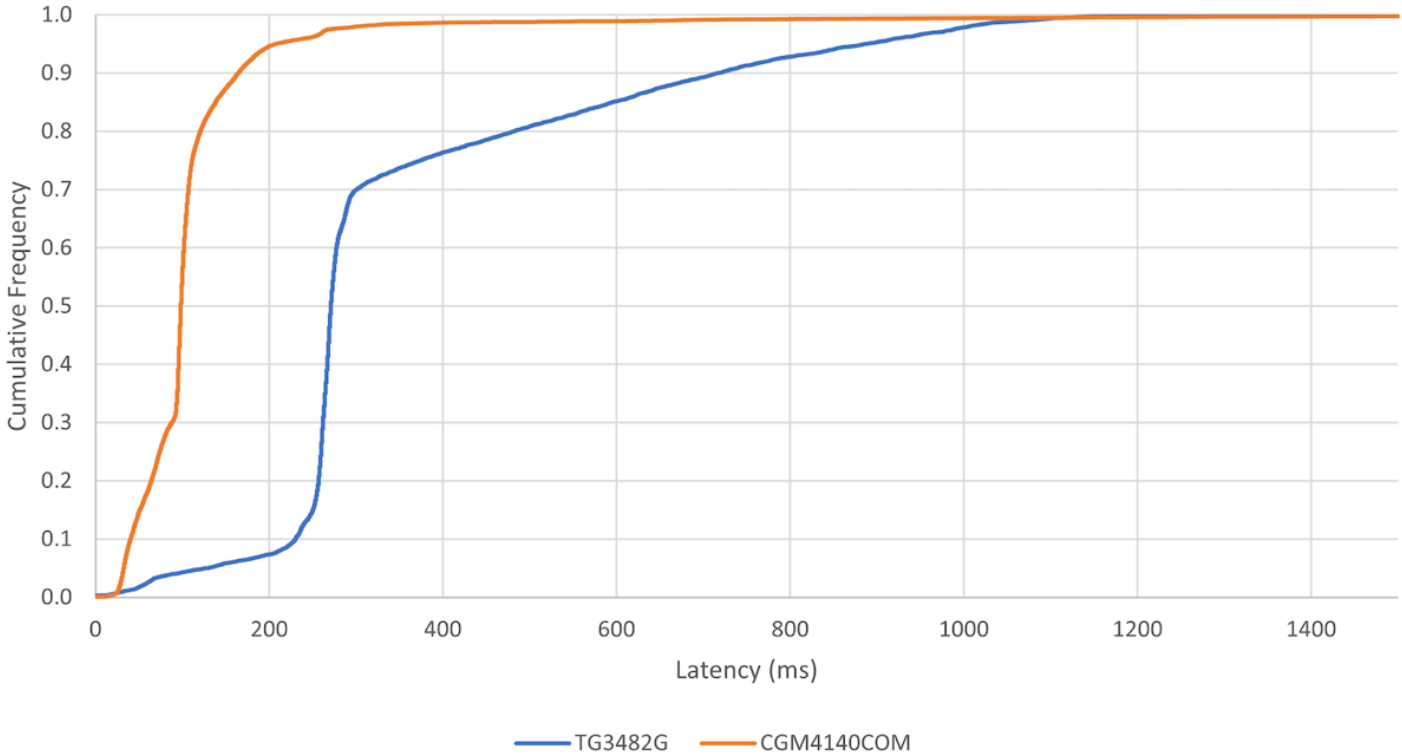

The following Figures 5 and 6 show the distribution of mean upstream latency for each of the two device variants under study. Figure 5 shows the variant without AQM, while figure 6 shows the variant that has AQM. The differences between the two distributions are quite interesting. In Figure 5 without AQM, there is a broad distribution from 15 milliseconds to several hundred milliseconds, with only 1% of results in the 15-30 millisecond range. In contrast, in Figure 6 with AQM, the distribution is tightly grouped between 15-45 milliseconds, with 77% of the results in the 15-30 millisecond range. As with the prior CDFs in Figures 4 and 5, this view of the data demonstrates not only substantially better latency performance with AQM but also much more consistent performance.

**Improving Latency with Active Queue Management (AQM) During COVID-19**

*Figure 5: Distribution of Mean Upstream Latency – CommScope CGM4140COM*

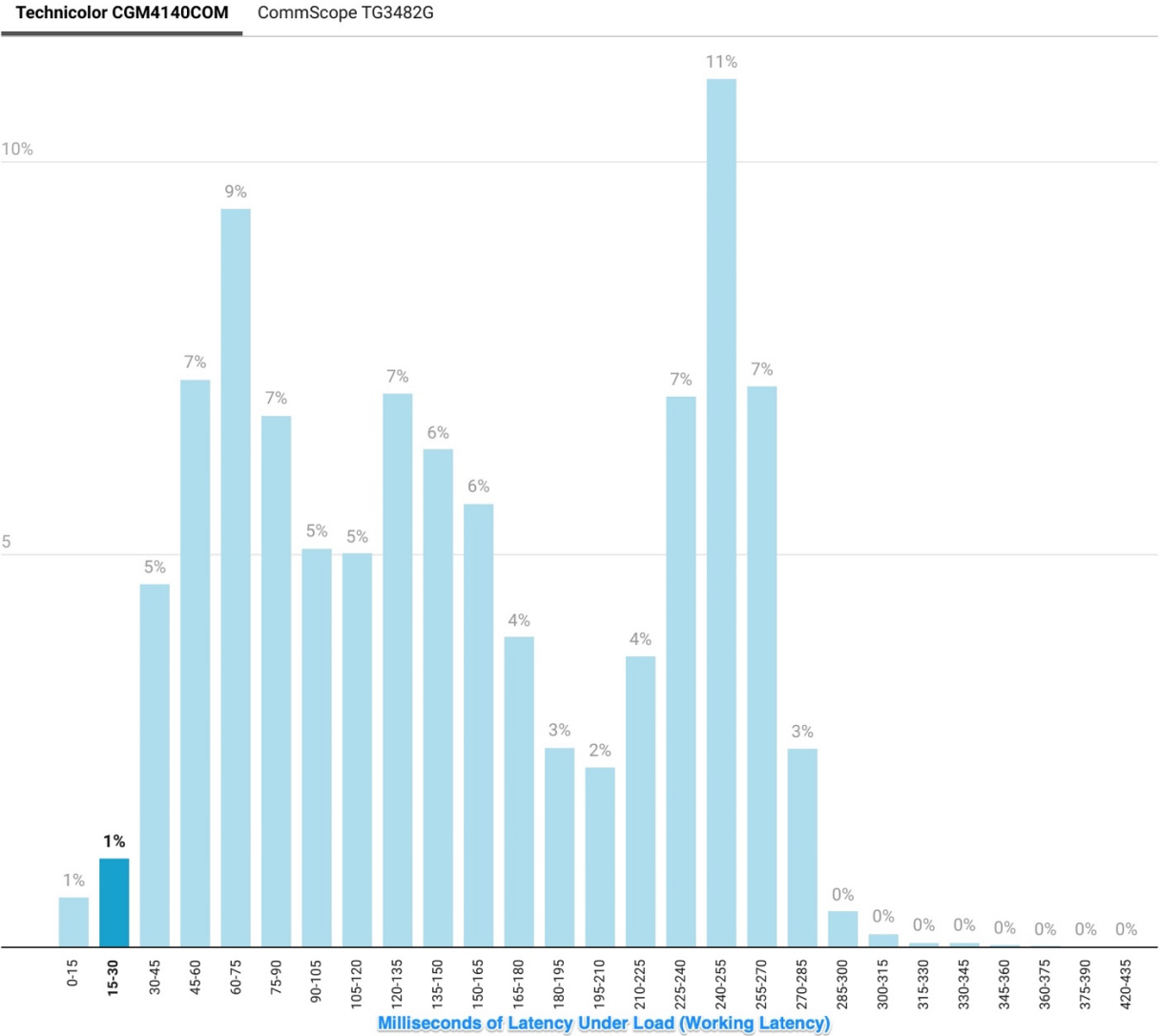

**Improving Latency with Active Queue Management (AQM) During COVID-19**

*Figure 6: Distribution of Mean Upstream Latency – Technicolor TG3482G*

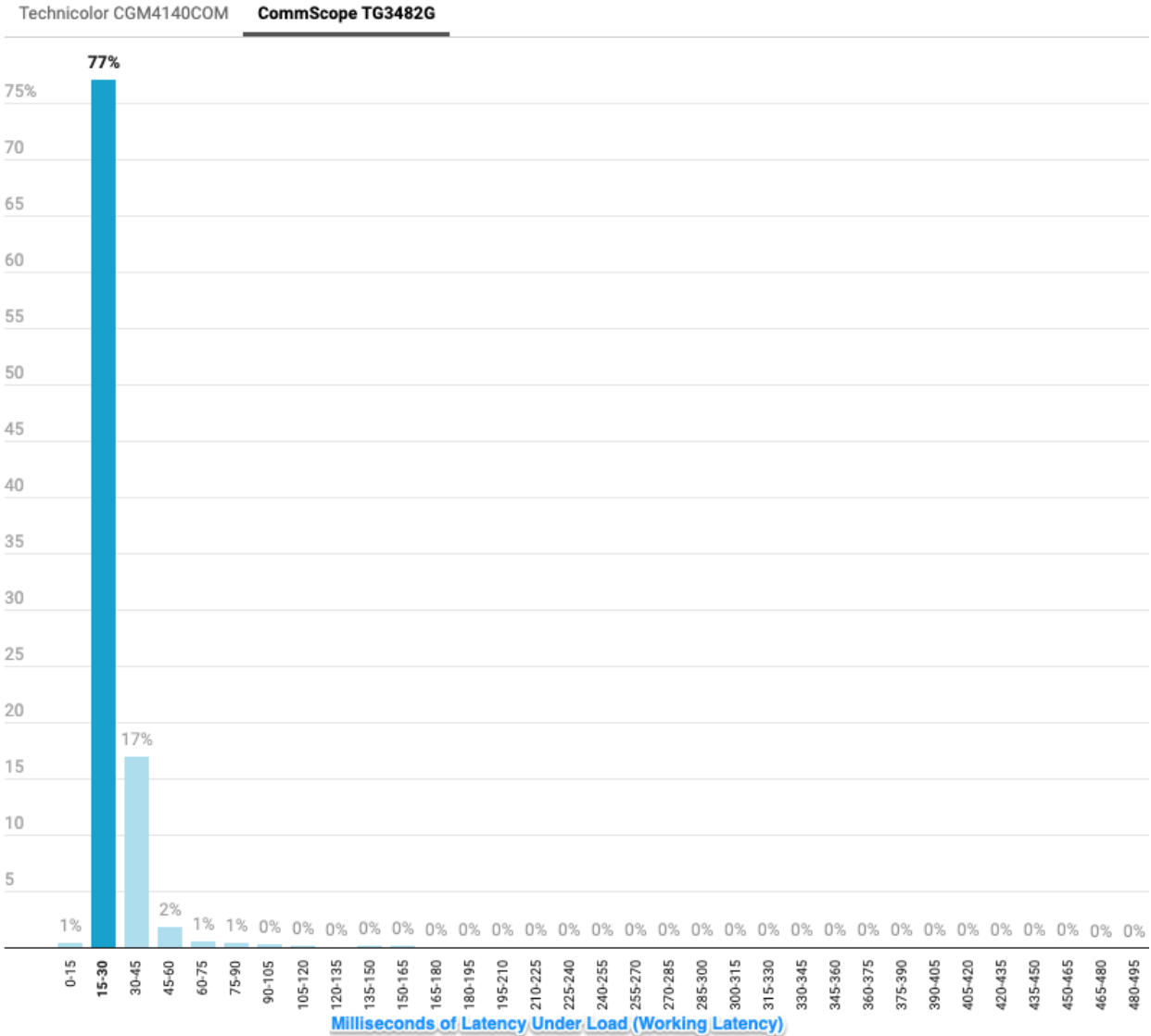

## 11. Conclusions

Just prior to the start of the pandemic, Comcast deployed a new Network Performance Measurement System that included a test for latency under load (working latency). That type of latency is central to an end user's Internet quality of experience (QoE), as became quite clear during the pandemic as a significant increase in video conferencing and other interactive Internet applications took place. Comcast moved quickly to optimize Active Queue Management (AQM) on the CMTS affecting downstream latency as a first step to ensure best QoE performance during COVID.

**Improving Latency with Active Queue Management (AQM) During COVID-19**

In addition, AQM functionality was deployed in cable modems in the Comcast network affecting upstream latency, though not at the same time on all CPE devices. That difference in deployment timing enabled us to study the performance of two variants of the same RDK-B-based cable modem gateway, one with and one without AQM for a period of time.

The data show that the latency under load performance differences between CPE devices with and without AQM is significant. At a time when users are more intensively using and depending upon latency-sensitive applications such as video conferencing, remote learning, and gaming, AQM is clearly a key tool to maximize their quality of experience. AQM also improves the user experience for all Internet applications where a human being is involved, such as web browsing, stock quotes, weather forecasts, maps, driving directions, etc. This improvement in QoE is also possible without any coordination with application developers or others, taking advantage of the loosely coupled architecture of the Internet and greatly simplifying any ISP and equipment manufacturer deployment of these performance improvements.

All Comcast DOCSIS 3.1 RDK-B-based gateway models have now been updated with DOCSIS-PIE AQM[59] and all are achieving dramatically improved working latency.

## 12. Next Steps

Latency under load is highly likely to emerge as an aspect of Internet service that is marketed to end users and which differentiates both ISPs and user-purchased CPE devices, particularly given end users' increasing adoption of latency sensitive applications. ISPs and CPE manufacturers should therefore continue to invest or start investing in AQM deployments, as Comcast continues to do, which will have an increasingly important and positive effect on end user performance now and in the future.

Beyond this type of AQM deployment, CableLabs has released the Low Latency DOCSIS (LLD) specifications for DOCSIS[60] [61]. This may significantly improve latency under load as compared to earlier implementations of AQM but is not yet fully implemented in cable modem and CMTS software deployed by Comcast and other DOCSIS-based ISP networks. Additionally, some of the enhancements rely on standardization work that is currently ongoing in the IETF[62]. But the prospect of reducing network delay to below 5 milliseconds at the 99th percentile could well enable the creation of entirely new classes of applications that were previously impossible or unimaginable, which is incredibly exciting for the future of the Internet.

## 13. Acknowledgements

The authors wish to thank the following individuals for reviewing drafts of this paper and providing valuable feedback: Bob Briscoe, Stuart Cheshire, Jim Partridge, Dan Rice, Dave Täht,

---

[59] AQM was enabled from launch in 2020 on the Technicolor XB7 (CGM4331COM). The CommScope XB7 (TG4482A) was fully enabled by February 1, 2021. AQM enablement for the TG3482G version of XB6 began the week of January 18, 2021, and completed by February 1, 2021.
[60] https://www.cablelabs.com/technologies/low-latency-docsis
[61] https://www.cablelabs.com/10g/latency
[62] Including assignment of a DiffServ code point and an Explicit Congestion Notification (ECN) codepoint.

**Improving Latency with Active Queue Management (AQM) During COVID-19**

Greg White, and Rich Woundy. We also wish to thank Aaron Tunstall for his contribution to the collection and analysis of the data, as well as Olakunle Ekundare for additional data to help contextualize these findings and Mulbah Dolley for performing analysis of related third-party datasets that helped to validate the latency trends we observed over the course of 2020.

*Document Update Log:*
*- 14 February 2022 – After publishing the first version on 30 July 2021, the document was updated to correct the labeling of Figures 5 and 6. The cable modem model numbers were correct, but the manufacturer names were transposed. In addition, the parenthetical was added to the X-axis to note that another name for latency under load was working latency.*